# Electronic and optical properties of C-N-codoped $TiO_2$: A first-principles GGA+U investigation

Guo MeiLi [a,b] and Du Jiulin [a]

[a] *Department of Physics, School of Science, Tianjin University, Tianjin 300072, People's Republic of China*

[b] *Department of Physics, School of Science, Tianjin Institute of Urban Construction, Tianjin 300384, People's Republic of China*

**Abstract**

Electronic structures and optical properties of C-N-codoped anatase $TiO_2$ were calculated by using GGA+U method based on the density functional theory. The calculated results showed that the N-doped, C-doped, and C-N-codoped $TiO_2$ produced $2p$ states in band gap, and the band gaps of the three doped systems decreased compared with the pure $TiO_2$. According to the optical results, the band edges of the three doped systems shifted to the long wavelength region, and the visible optical absorption from 450 to 800 nm was observed. Moreover, the visible light response of C-N-codoped $TiO_2$ was better than the C or N single doped $TiO_2$, indicating that there was a synergistic effect for the C-N-codoped $TiO_2$, which offseted the deficiencies of C or N-doped $TiO_2$.

**1. Introduction**

$TiO_2$ has been extensively studied as the most promising photocatalyst in hydrogen production and environmental protection fields because of its high efficiency, low cost, nontoxcity and photostability. Unfortunately, anatase $TiO_2$ has a wide band gap (about 3.2 eV), which only responds to UV light irradiation accounting for only a small fraction of solar light, while visible light occupying most fraction of solar light can't be utilized. Therefore, there have been many methods to be used to broaden the photoresponse of $TiO_2$ to the visible light region, such as metal and nonmetal doping [1-5], semiconductor compounding [6, 7], and dye sensitizing [8, 9], among which, the nonmetal element doping is considered as one of the most efficient methods [10-13].

Among the nonmetal elements, N has been widely investigated both experimentally and theoretically to improve the visible absorption of $TiO_2$. Asahi *et al.* calculated the band structure and optical property of N-doped $TiO_2$, and observed a red shift of absorption edge. According to their results, the substitution doping of N could lead to the band-gap narrowing by mixing its $2p$ states with O $2p$ states [1]. Meanwhile, they prepared films and powders of N-doped $TiO_2$, and revealed a dramatic improvement in optical absorption and the level of photocatalytic activity under visible light compared with the pure TiO$_2$. Sathish *et al.* synthesized N-doped TiO$_2$ nanocatalyst, and the results showed that after N doping, the light absorption shifted to 550 nm [14].

More recently, N and nonmetal element codoped TiO$_2$ has attracted considerable interests, since it can result in a higher visible photocatalytic activity compared with the single N-doped TiO$_2$. It has been reported that S-N, F-N, B-N, and C-N-codoped TiO$_2$ can significantly improve the photocatalytic efficiency under visible light illumination [15-19]. For the C-N-codoped TiO$_2$, Zhang *et al.* confirmed that the C-N-codoping could give rise to a synergistic effect [20]. They speculated that C dopant's states could connect with N dopants' states, therefore, leading to a band gap narrowing. Indeed, their experiment results verified the speculation. The results of Yang *et al.* indicated that N atom could incorporate into the lattice of anatase TiO$_2$ through substituting the sites of O atoms, while most of C atoms could form a mixed layer of deposited active carbon and complex carbonate species at the surface of TiO$_2$ nanoparticles [21, 22]. Carbonaceous species formed by doped C atoms acted a role of photosensitizer like organic dyes, and could be excited and inject electrons into the conduction band of TiO$_2$, accordingly, improving photocatalytic activity. Lim *et al.* also synthesized the C-N-codoped TiO$_2$, which exhibited a higher photocatalytic activity under visible light irradiation than pure, C-doped, and N-doped TiO$_2$ [23-26].

However, different experimental conditions and sample preparation methods make it difficult to understand codoping synergistic mechanism. Computer simulation can overcome complexity of the experimental conditions and help us to analyze the microscopic information of electronic structure of C-N-codoped TiO$_2$, and to understand how the N and C dopant states impact the visible light absorption of TiO$_2$.

In this work，using the GGA+U method based on the density functional theory (DFT), we studied the electronic structures and optical properties of N-doped, C-doped, and C-N-codoped TiO$_2$, respectively. Compared with bare GGA, GGA+U can correct the underestimated band gap. The synergistic effect of C and N codoping was discussed. The optical transition mechanisms of N-doped, C-doped, and C-N-codoped TiO$_2$ were also proposed, respectively.

**2. Calculation methods**

We performed the DFT calculations using Cambridge Serial Total Energy Package (CASTEP) [27], within the plane-wave-pseudo-potential approach [28], together with the Perdew-Burke-Ernzerhof (PBE) exchange correlation functional [29]. The interaction between valence electrons and the ionic core was described by ultrasoft pseudopotential [30, 31], which was used with $2s^22p^4$, $3d^24s^2$, $2s^22p^2$ and $2s^22$ $^3$ as the valence electrons configuration for the O, Ti, C, and N atoms, respectively. We chose the energy cutoff to be 340 eV for the pure, N-doped, C-doped, and C-N-codoped TiO$_2$. The Brillouin-zone sampling mesh parameters for the k-point set were 3×2×3 [32, 33]. And then, Geometry optimization was performed by GGA method, because the GGA calculation on the geometry parameters agree better with the experiments than the GGA+U method [29, 34]. In this way, atomic positions and lattice parameters were optimized until the atomic forces were below 0.01 eV/°A. The GGA+U approach introduces an intra-atomic electron-electron interaction as an on-site correction in order to describe systems with localized *d* and *f* states, which can produce better band gap relative to GGA. To account for the strongly correlated interactions of the Ti 3*d* electrons, a moderate on-site Coulomb repulsion U=6.6 eV was applied to the further calculations of electronic structures and optical properties after careful checking the U dependent band gap [35].

To calculate the electronic structures and optical properties of pure, N-doped, C-doped, and C-N-codoped $TiO_2$, $2 \times 3 \times 1$ anatase supercell was used, which contained 72 atoms. The substitution method was taken into account in this paper. The structure of C or N-doped anatase $TiO_2$ was modeled by replacing one O atom with one C or N atom. For the C-N-codoped anatase $TiO_2$, two O atoms were replaced with one C and one N atom. The recent investigation indicated that the covalent bond between C atom and adjacent O atom could be enhanced easily when the distance between C and N atoms was small, and the covalent bonds made the carrier transfer more difficult. When the distance between C and N atoms became larger, their interactions could be reduced, which was beneficial to electrons transition [36]. Therefore, the distance between C and N atoms in the C-N-codoped $TiO_2$ should be as far as possible. In this case, the doped $TiO_2$ systems formed the configurations of $Ti_{24}O_{47}C$, $Ti_{24}O_{47}N$, and $Ti_{24}O_{46}CN$, corresponding to the concentration of 2.08%. The calculated structures of substitution doping are presented in Fig. 1. After optimized, the lattice constants have minor changes.

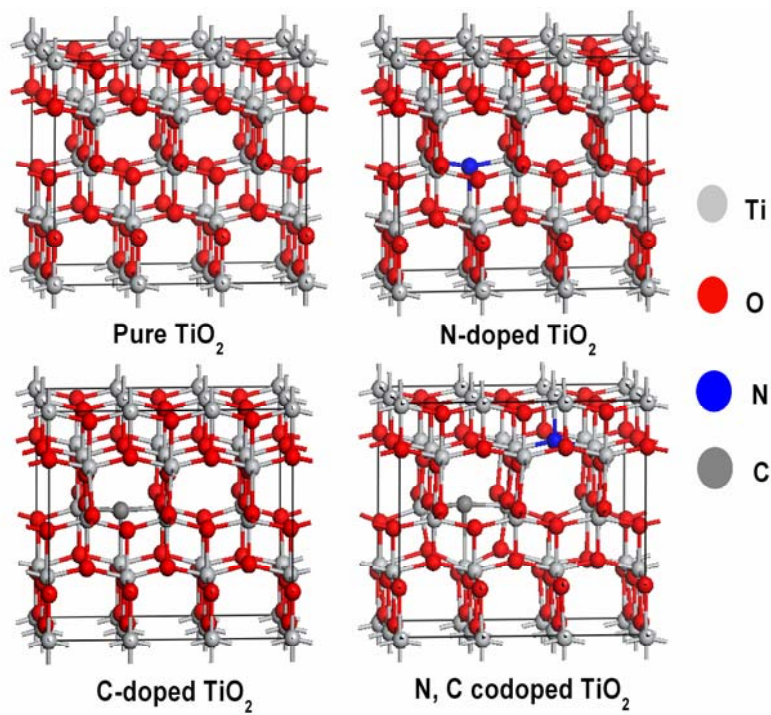

Figure 1 Optimized crystal structures of pure, N-doped, C-doped, and C-N-codoped $TiO_2$.

## 3. Results and discussions
### 3.1. Band structure and DOS of pure and doped anatase $TiO_2$

The band structure of pure anatase $TiO_2$ has been calculated by the GGA and GGA+U methods, respectively. It is observed in Fig. 2a that the band gap calculated by the GGA method is about 2.14 eV, which is consistent with previous theoretical studies [37, 38], but is much less than 3.2 eV of the experimental result. It is well known that the underestimated band gap can be due to the choice of exchange-correlation energy. The failure of GGA can be associated with an inadequate description of the strong Coulomb interaction between $3d$ electrons localized on Ti atoms. The underestimation of band gap can be corrected by introducing an additional term U, which takes into account the repulsion between two electrons placed on the same 3*d* orbital. To overcome this deficiency, we have carried out the GGA+U calculations with different U values for Ti 3*d* states and then, the band gap has been corrected to about 2.89 eV for pure anatase $TiO_2$ for U=6.6 eV (Fig. 2b), better consistent with experiment value of 3.2 eV [39]. We choose U=6.6 eV, because the U value is used in previous works and it is conceived that it is reliable [40]. For the band structure of pure $TiO_2$, it can be seen that the top of valence band and the bottom of the conduction band locate at a different position, which means that $TiO_2$ with anatase structure is an indirect-gap material. The valence band of pure $TiO_2$ mainly consists of the 2*p*, 2*s* states of O and 3*d* states of Ti. In the valence band, the O 2*p* states are mainly between −5 and 0 eV, while the O 2*s* states appear in the range from −18 to −15.5 eV. The Ti 3*d* states give rise to some bands in the energy range from −5 to −3 eV. The lowest conduction band is dominated by Ti 3*d* states.

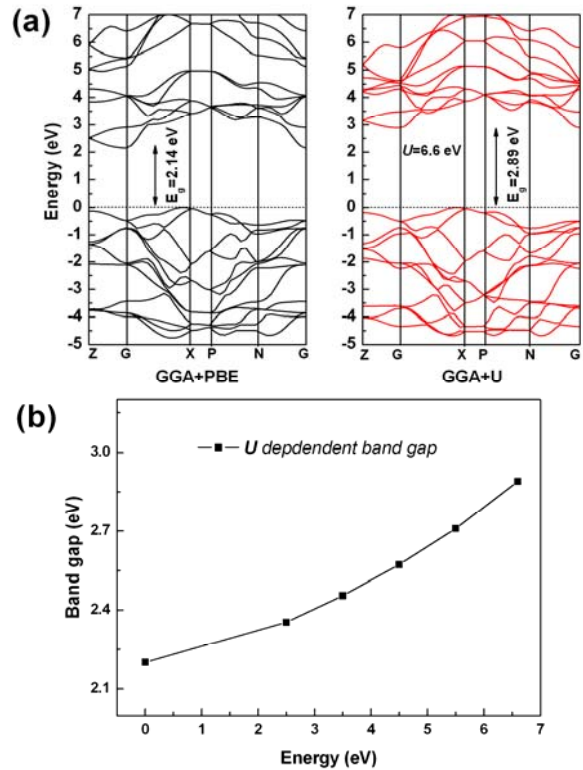

Figure 2 (a) Band structures of pure $TiO_2$ based on GGA and GGA+U methods, (b) U value dependent band gap.

To explore the electronic structures of N-doped, C-doped, and C-N-codoped $TiO_2$, we calculate density of states (DOS), which are presented in Fig. 3. Compared with pure $TiO_2$, the C doping can induce the mid-gap states in both the spin-up and the spin-down bands, which mainly consist of the C 2*p* electronic states. These mid-gap states are symmetric. N doping introduces band gap states of the N 2*p* in the spin-down bands, however, no band gap states are observed in the spin-up bands, which are similar to the previous result [41]. From the results of DOS of C-N-codoped $TiO_2$, we can also observe the mid-gap states in both the spin-up and the spin-down bands, and these states are also asymmetric due to the synergistic effect of N and C codoping. In Fig. 3, we can observe that the band gaps of C-doped, N-doped, and C-N-codoped TiO$_2$ are decreased to 2.76, 2.87, and 2.74 eV, respectively, which is consistent with the previous work [23]. Compared with metal and nonmetal codoped system, the band gap narrowing is similar. For example, it is reported that C-W and C-Mo codoped can decrease the band gap by introducing additional dopant states, and have been seen as effective visible-light photocatalysts [42, 43]. Inspired by electronic structure, it is interesting to further investigate the optical properties.

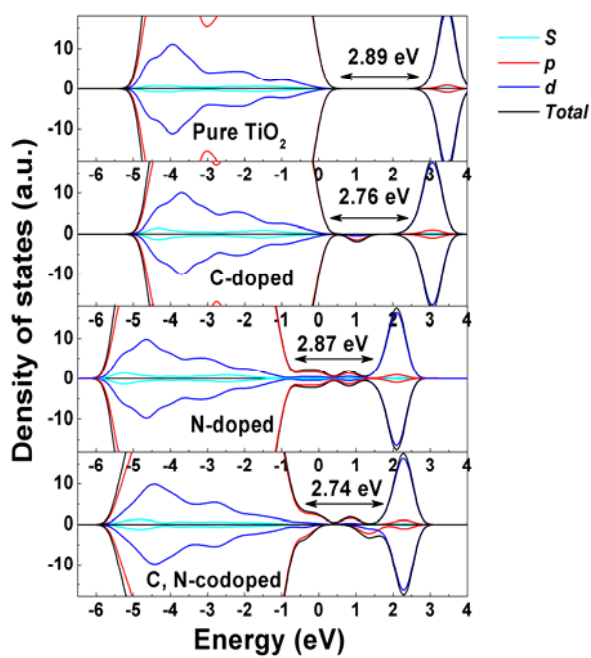

Figure 3 The DOS of pure, N-doped, C-doped, and C-N-codoped TiO$_2$.

## 3.2. Optical properties of pure and doped anatase TiO$_2$

To investigate the optical band gap and optical transition of C-N-codoped TiO$_2$, it is necessary to investigate the imaginary part of the dielectric function $\varepsilon_2(\omega)$, because $\varepsilon_2(\omega)$ is important for describing the optical properties of any materials [44, 45]. It is well known that the interaction of a photon with electrons in a material can be described in terms of time-dependent perturbations of the ground-state electronic states. Optical transitions between occupied and unoccupied states are caused by the electric field of the photon. The spectra from the excited states can be described as a joint density of states between the valence and conduction band. The momentum matrix elements, which are used to calculate $\varepsilon_2(\omega)$, are calculated between occupied and unoccupied states, which are given by the eigenvectors obtained as solution of the corresponding Schrödinger equation. To evaluate these matrix elements, one uses the corresponding eigenfunctions of each of the occupied and unoccupied states.

Figure 4 gives the imaginary part of the dielectric function $\varepsilon_2(\omega)$ of pure, N-doped, C-doped, and C-N-codoped TiO$_2$. The dielectric function spectrum is important indicator for any materials to describe the optical response. Optical transitions between occupied and unoccupied states are caused by the electric field of the photon. The spectra from the excited states can be described as a joint DOS between the valence and conduction bands. Optical transition peaks correspond to optical transitions between two states, and intensity of peaks is proportional to density of states. For the pure $TiO_2$, with increase of energy response, 4.29 eV optical transition can be observed, which indicate optical response of the imaginary part of the dielectric function for band gap. After C and N incorporation, the optical transition from band gap decreases to 3.91 eV for C-doped, 4.22 eV for N-doped, and 3.86 eV for C-N-codoped $TiO_2$, respectively. The redshift of the optical transition indicates that the band gaps of the three doped systems are decreased, which is in good agreement with the result of DOS. Further, the optical transitions in the band gap have been investigated. 1.21 eV optical transition peak is observed in the low energy region for the C-doped $TiO_2$, which may be related to the transition of valence band to dopant states. Meanwhile, 2 eV optical transition peak can be observed for the N and C-N codoped $TiO_2$ systems, which indicate that localized N and/or C $2p$ states have shown visible optical transition. Therefore, it is necessary to investigate visible absorptions of the three doped systems by considering the band gap narrowing and localized hybrid *p* states.

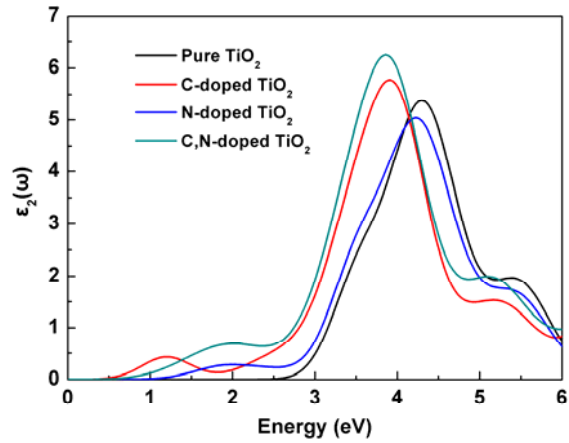

Figure 4 The imaginary part of dielectric function $\varepsilon_2(\omega)$ of pure, N-doped, C-doped, and C-N-codoped $TiO_2$.

Figure 5 presents the optical absorptions of pure and three doped systems. It is shown that the optical band edges of C-doped, N-doped, and C-N-codoped $TiO_2$ shift obviously to the visible light region, which is consistent with the electronic structure calculation. Further, it is found that the calculated optical band gaps decrease from 3.3 eV for pure $TiO_2$ to 2.89 eV for C-doped, 2.98 eV for N-doped, and 2.91 eV for C-N-codoped $TiO_2$, respectively, which are good consistent with the experimental data [20, 24, 46]. Meanwhile, the optical absorptions between 450-800 nm are enhanced for the three doped systems compared with the pure $TiO_2$. Especially, the absorption intensity of C-N-codoped $TiO_2$ is obviously higher than that of single N or C-doped $TiO_2$. Lin *et al* calculated the optical properties of N-doped $TiO_2$ by using LDA method, and found visible absorption in the range of 400-500 nm, while the optical absorption center of oxygen-deficient $TiO_2$ was located at 500 nm[47]. It was also found that the C-doped $TiO_2$ could form the localized states and thus induced the significant visible optical absorption [48-50]. Besides, the recent investigation also showed the N-C codoped $TiO_2$ could present the stronger visible light absorption in the range of 400-600 nm than that of C or N single doped $TiO_2$[51]. Our work used

GGA+U method to correct band gap and clearly showed that the visible optical absorption of C-N codoped TiO$_2$ should be reliable.

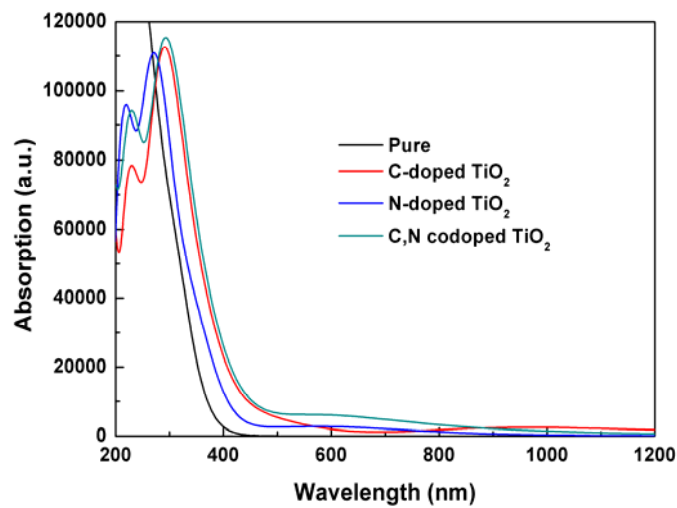

Figure 5 The optical absorptions of pure, N-doped, C-doped, and C-N-codoped TiO$_2$.

Figure 6 gives the band structures of three doped systems. According to the results, for the C-doped TiO$_2$, three bands of the localized 2*p* states, locating at around -0.75, -0.31, and 1.2 eV, appear in the band gap. The band at -0.75 eV behave as a part of valance band and contribute to the band gap narrowing, while the other bands (-0.31 and 1.2 eV band) lead to optical absorption in the region of 500-800 nm. Similarly, the N-doped TiO$_2$ can also produce N 2*p* states near valence band and middle states in the band gap, and correspond to contribution of band gap narrowing and visible optical absorption, respectively. Previous work showed that the N-doped TiO$_2$ was similar with the C-doped TiO$_2$, where N 2*p* or C 2*p* states behave as the localized electronic states near the valence band and middle states in the gap[49]. Our work is in good agreement with it. After N and C codoping, the band edge of TiO$_2$ present remarkable shift to visible range compared with single doping, and wide optical absorption with centre of 600 nm is obviously observed. Indeed, the C-doped TiO$_2$ induces more significant band gap narrowing while the N-doped TiO$_2$ lead to stronger visible optical absorption, and thus C-N codoped induces both of significant band gap narrowing and visible optical absorption. The visible absorption of TiO$_2$ is enhanced due to the synergistic effect. Combined with optical absorption in Fig. 5, the enhanced optical absorption can be observed in 450-800 nm, which indicate that the hybrid *p* states formed by C 2*p* and N 2*p* states have dominant contribution to visible optical absorption. Therefore, combined with the theory and the experimental results, it is shown that C-N-codoping can further enhance visible light absorption due to the synergistic effect of N and C 2*p* states hybridization.

## 4. Conclusion

In summary, the electronic structures and optical properties of C-N-codoped TiO$_2$ have been evaluated on the basis of GGA+U calculations. In the cases of C- and N-doped TiO$_2$, the doped C and N atoms introduced 2*p* states, which induced the band gap narrowing and visible absorption. In the C-N codoped system, we also observed the band gap states of 2*p*, which were composed of C 2*p* and N 2*p* states. Meanwhile, the band gap of C-N-codoped TiO$_2$ was smaller than that of N and C single doped $TiO_2$. From the optical properties of C-N-codoped $TiO_2$, we observed the absorption enhancement in the range of 450-800 nm, which was ascribed to the transition of N and/or C $2p$-Ti $3d$ states. The visible absorption of C-N-codoped $TiO_2$ was better than that of C or N single doped $TiO_2$ due to the synergistic effect.

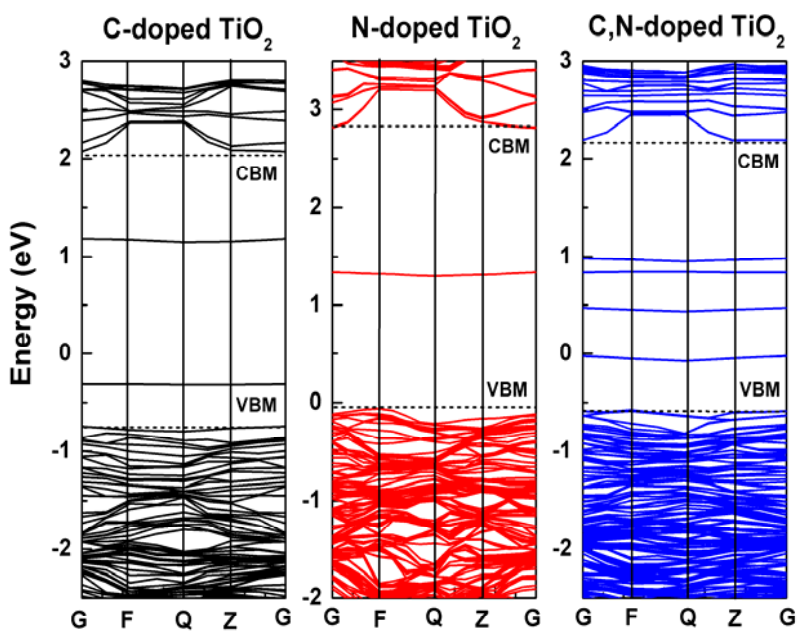

Figure 6 The band structures of C-doped, N-doped, C-N codoped $TiO_2$. CBM and VBM represent conduction band minimum and valence band maximum, respectively

**Acknowledgments**

Guo ML hopes to thank Prof. Yanni Li, School of Chemical Engineering and Technology at Tianjin University, for providing us with the computational plat. This work is supported by the National Natural Science Foundation of China (Grant No. 11175128) and by the Higher School Specialized Research Fund for Doctoral Program (Grant No. 20110032110058).